# Dynamic Disorder in Negative Thermal Expansion Compound Zn(CN)$_2$

R. Mittal[1,2*], S. Mitra[1], H. Schober[3], S. L. Chaplot[1] and R. Mukhopadhyay[1]

[1]Solid State Physics Division, Bhabha Atomic Research Centre, Mumbai, India
[2]Juelich Centre for Neutron Science, IFF, Forschungszentrum Juelich, Garching, Germany
[3]Institut Laue-Langevin, BP 156 X, F-38042 Grenoble Cedex, France

**Abstract**

Dynamical disorder in negative thermal expansion compound Zn(CN)$_2$ is investigated by quasielastic neutron scattering technique in the temperature range 170-320 K. Significant quasielastic broadening is observed above the phase transition temperature of ~250 K, however no broadening is observed at 220 K and below. Data at high temperatures are analyzed assuming the CN reorientation. Characteristic time associated with the CN orientation is estimated as 16 ps and 11 ps at 270 and 320 K respectively.

**Introduction**:

The materials exhibiting unusual properties like negative thermal expansion (NTE) are of interest due to both fundamental scientific importance and potential applications [1-3]. These materials find applications in various fields e.g., in making composites with nearly zero thermal expansion coefficient by compensating with the usual positive thermal expansion coefficient of other materials. Oxide based framework materials are perhaps the most widely studied NTE materials [1,4,5]. Besides oxide based framework materials, NTE behaviour has been also observed in molecular framework materials containing linear diatomic bridges such as the cyanide anion. One such material is Zn(CN)$_2$ which is reported to have a NTE coefficient twice as large as that of ZrW$_2$O$_8$ [2].

NTE phenomenon is known to arise from a range of different physical mechanisms. For example, magnetostriction in ferroelectric materials [5], valence transitions in intermetallic [6] and fulleride [7] materials and the population of low energy phonon modes in oxide-based framework materials. In oxide based materials, it can be considered that there exists a network of coordination polyhedra, connected via 1) single atom metal –oxygen-metal (M-O-M') or 2) oxygen-metal-oxygen (O-M-O'). Examples are Zr-O-W linkage in ZrW$_2$O$_8$ and O-Cu-O linkage in Cu$_2$O. It is now accepted that the mechanism driving NTE in these materials involves transverse vibrational displacement of the central linking atom. With increase in temperature, the central linking atom goes away from the M...M' or O..O' axes which in turn draws the two anchoring atoms closer together and thereby giving a NTE phenomenon.

Structural studies showed that two different models (one ordered and one disordered) fit equally well to the diffraction data and give similar agreement factor. Neutron diffraction studies [8] have shown that the Zn(CN)$_2$ has cubic symmetry with space group *pn3m* with *a* = 5.9227 Å at 14 K and 5.8917 Å at 305 K. The CN groups are found to be disordered in contrast to the ordered arrangement with symmetry *p43m* assumed in earlier X-ray diffraction experiments [9]. The structure of $Zn(CN)_2$ is shown in Fig. 1. Therefore, $Zn(CN)_2$ type of materials are fundamentally different than oxide-based framework materials described above in a way that their structures are based on a framework of metal-cyanide-metal (M-CN-M') type rather than M-O-M' type. Atomic pair distribution function (PDF) analysis [10] of high energy X-ray scattering data showed an increase of the average transverse thermal amplitude of the bridging C/N atoms away from the body diagonal with heating from 100 to 400 K. This increase of the thermal amplitude of the bridging atoms is believed to be the origin of the NTE behaviour in $Zn(CN)_2$.

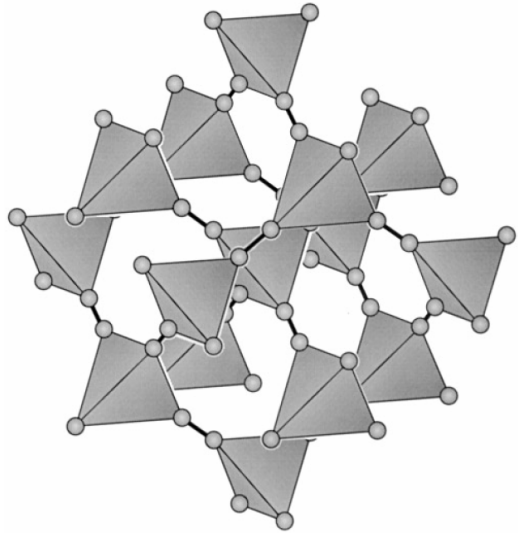

Fig. 1 The structure of $Zn(CN)_2$ [8]. The tetrahedra of four (C, N) atoms (spheres) are shown connected by heavy lines representing C-N bonds.

Differential Scanning Calorimetry (DSC) study (Fig. 2) also showed a second order phase transition in $Zn(CN)_2$ around 250 K. It is important, therefore, to study the disordered nature of CN as a function of temperature in order to understand the NTE phenomenon in $Zn(CN)_2$ material in detail. The compounds NaCN, KCN and RbCN [11] also show order-disorder transition related to CN ion orientation at around 200 K. However, no quasi-elastic neutron scattering studies have been reported in these compounds. The temperature range of CN order-disorder transition is similar to as found in DSC measurements of $Zn(CN)_2$.

The large negative Grüneisen parameters of low energy phonon modes are responsible [12,13] for the negative thermal expansion in these materials. For $Zn(CN)_2$, temperature dependence of Grüneisen parameter of low energy phonons have been recently determined [14] using high pressure inelastic neutron scattering measurements. The Grüneisen parameter values show a substantial change at 275 K compared to the values estimated from the experimental data at 165 K and 225 K. The change in Grüneisen parameters values at 275 K is believed to be due to the onset of dynamical disorder of CN orientation.

In this paper, the results of quasielastic neutron scattering study on Zn(CN)$_2$ compound at different temperatures above and below the phase transition temperature of 250 K are presented.

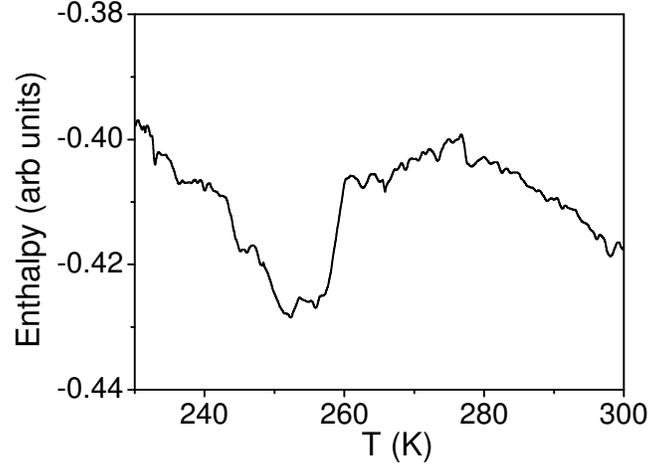

Fig. 2 The enthalpy measurements for Zn(CN)$_2$ as a function of temperature.

**Experiment:**

Powder Zn(CN)$_2$ (98% purity) sample was procured commercially from Sigma-Aldrich and used as it is. Quasielastic coherent neutron scattering measurement was performed using the IN6 spectrometer at the Institut Laue-Langevin (Grenoble, France). IN6 is a time-focussing time-of-flight spectrometer designed for quasielastic and inelastic scattering for incident wavelengths in the range of 4 to 6 Å. For the present study, incident wavelength of 5.1 Å was used which provided an elastic energy resolution of 80 μeV. Data was recorded at temperatures below and above phase transition temperature (~250 K) in the range of 170-320 K. Thickness of sample container was chosen such a way that the multiple scattering has a negligible effect on the QENS data.

**Results and Discussion**:

In a neutron scattering experiment the measured intensity is proportional to the double differential scattering cross section, which in turn is proportional to the scattering law S(**Q**,ω). Here **Q** (= **k**$_f$ - **k**$_i$) is the wavevector transfer, **k**$_f$ and **k**$_i$ being the magnitudes of the wavevector of the neutrons after and before the scattering respectively, and ω is the angular frequency corresponding to the energy transfer, ℏω= E$_f$ - E$_i$, E$_i$ and E$_f$ being the initial and final energies respectively of the neutrons. According to Van Hove, the coherent and incoherent double-differential scattering cross sections and the scattering laws are given by [15]

$$\left[\frac{d^2\sigma}{d\Omega d\omega}\right]_{inc} = \frac{1}{N}\frac{k_f}{k_i}\sum_{\alpha=1}^{n} b_\alpha^{inc} S_{inc}^\alpha(Q,\omega) \quad (1)$$

where

$$S_{inc}^\alpha(\boldsymbol{Q},\omega) = \frac{1}{2\pi N_\alpha}\int_{-\infty}^{\infty}\sum_{i_\alpha=1}^{N}\left\langle \exp\{i\boldsymbol{Q}.\boldsymbol{R}_{i_\alpha}(t)\}\exp\{-i\boldsymbol{Q}.\boldsymbol{R}_{i_\alpha}(0)\}\right\rangle \exp(-i\omega t)dt \quad (2)$$

And

$$\left[\frac{\partial^2 \sigma}{\partial \Omega \partial \omega}\right]_{coh} = \frac{k_f}{k_i}\frac{1}{N}\sum_{\alpha}^{n}\sum_{\beta}^{n} b_\alpha^{coh} b_\beta^{coh} \sqrt{N_\alpha N_\beta} S_{coh}^{\alpha\beta}(\boldsymbol{Q},\omega) \quad (3)$$

where

$$S_{coh}^{\alpha\beta}(\boldsymbol{Q},\omega) = \frac{1}{2\pi\sqrt{N_\alpha N_\beta}}\int_{-\infty}^{\infty}\sum_{i_\alpha=1}^{N_\alpha}\sum_{j_\beta=1}^{N_\beta}\left\langle \exp\{i\boldsymbol{Q}.\boldsymbol{R}_{i_\alpha}(t)\}\exp\{-i\boldsymbol{Q}.\boldsymbol{R}_{j_\beta}(0)\}\right\rangle \exp(-i\omega t)dt \quad (4)$$

Here $b_\alpha^{inc}$ is the incoherent scattering length for the species α and $b_\alpha^{coh}, b_\beta^{coh}$ are the coherent scattering lengths relative to species α and β respectively. A comparison between Eq. (2) and (4) reveals that in the case of incoherent scattering one must keep track of the history of one single particle while in the case of coherent scattering one has to follow up the simultaneous evolution of all particles. In case of rotation, the general incoherent scattering law, $S_{inc}(\boldsymbol{Q},\omega)$ consists of an elastic component and a quasielastic component. However, in case of coherent scattering, no elastic component is expected in the rotational scattering law, $S_{coh}(\boldsymbol{Q},\omega)$, if the measurements are done away from the Bragg peaks. However, coherent scattering law will consist of an elastic component even away from the Bragg peak if there is any static disorder present in the system as discussed in Ref. 16.

Typical quasielastic spectra at Q= 0.9 A$^{-1}$ from Zn(CN)$_2$ are shown in Fig. 3 at different temperatures. It can be clearly seen from Fig. 3 that below the phase transition (at 170 and 220 K), data do not show any quasielastic broadening. However, quasielastic broadening is observed at 270 and 320 K, which are above the transition temperature of 240 K. Therefore, it is considered that the broadening above the phase transition temperature is due to dynamical disorder of CN orientations since the structural studies [8] indicated that the CN bonds are orientationally disordered or flipped randomly in its disordered phase.

It is clear from the QENS spectra shown in Fig. 3 that both elastic and quasielastic components are present in the spectra taken above the transition temperature. Origin of this elastic component could be due to the presence of static disorder. The disordered orientations are known to be frozen at low temperature and are likely to be partially frozen above the transition at 240 K. Therefore, data was analysed by assuming the following scattering law

$$S_{coh}(Q,\omega) = A(Q)\delta(\omega) + B(Q)L(\omega,\Gamma(Q)) \quad (5)$$

Here the first term represents the elastic component and second term is the quasielastic component which is approximated with a Lorentzian function of half width at half maxima (HWHM), Γ(Q). It was found that Eq. (5) fitted very well with the data at 270 and 320 K. Fig. 4 shows the typical spectra fitted with Eq. (5) at T=320 K.

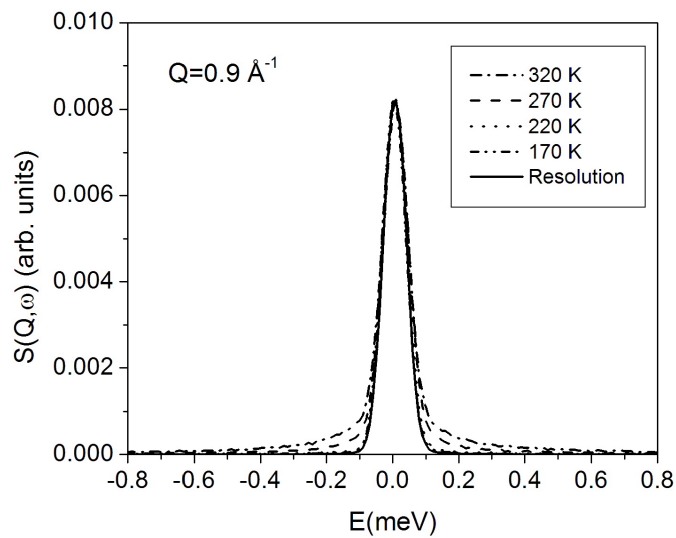

Fig. 3 Typical quasielastic spectra at Q=0.9 Å$^{-1}$ at different temperatures. Resolution of the spectrometer is shown by the solid line.

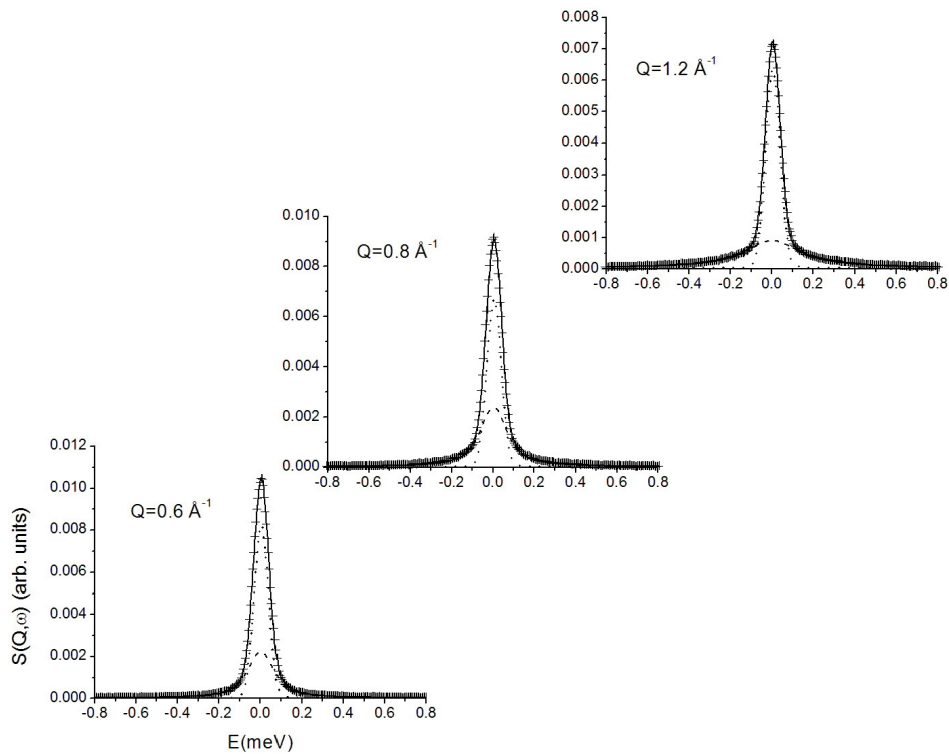

Fig. 4 Typical quasielastic spectra for few Q values at 320 K. Dashed line is the quasielastic component and dotted line is the elastic component.

To determine the nature of reorientation of CN bond, it is better to plot the variation of quasielastic structure factor with Q and then compare that with the possible model. The variation of quasielastic structure factor, B(Q) is shown in Fig. 5. Behaviour of the extracted quasielastic structure factor with Q is found to be almost unchanged up to the measured range of Q=2 Å$^{-1}$ as expected for the CN flipping motion. However, to confirm this assertion, data with Q-values up to 4 Å$^{-1}$ would be required considering the distance between C and N atoms (1.18 Å). The residence time is obtained from the width of the quasielastic component. The resident time in the disordered state has been found to be ~16 ps and 11 ps at 270 and 320 K respectively.

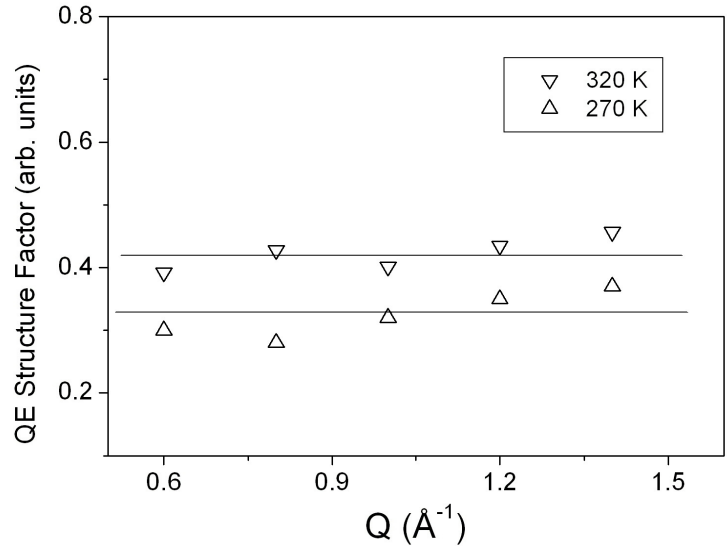

Fig. 5 Variation of quasielastic structure factor (B(Q) in Eq. (5) ) for different temperatures.

**Conclusions**:

Coherent quasielastic neutron scattering data are analysed in terms of CN reorientation/flipping in technologically important $Zn(CN)_2$ compound which shows negative thermal expansion behaviour in wide range of temperatures. Significant quasielastic broadening is observed above the phase transition temperature (~250 K) while below the phase transition temperature, very small or no quasielastic broadening is observed. This clearly establishes the dynamic disordered nature of CN bond above the phase transition. Approximate resident times in the disordered state has been found to be ~16 ps at 270 K and 11 ps at 320 K.